\documentclass[acmsmall]{acmart}
\usepackage[frozencache=true, cachedir=_minted-output]{minted}
\usepackage{todonotes}
\setuptodonotes{inline}

\AtBeginDocument{%
  \providecommand\BibTeX{{%
    \normalfont B\kern-0.5em{\scshape i\kern-0.25em b}\kern-0.8em\TeX}}}

\acmYear{2020}

\acmVolume{}
\acmNumber{}
\acmArticle{5}
\acmMonth{11}

\setcopyright{none}
\renewcommand\footnotetextcopyrightpermission[1]{}

\begin{document}

\title{Opportunities and Challenges for Circuit Board Level Hardware Description Languages}


\author{Richard Lin}
\email{richard.lin@berkeley.edu}
\affiliation{%
  \institution{University of California, Berkeley}
}

\author{Bj\"orn Hartmann}
\email{bjoern@berkeley.edu}
\affiliation{%
  \institution{University of California, Berkeley}
}

\renewcommand{\shortauthors}{Lin, et al.}

\begin{abstract}

Board-level hardware description languages (HDLs) are one approach to increasing automation and raising the level of abstraction for designing electronics.
These systems borrow programming languages concepts like generators and type systems, but also must be designed with human factors in mind to serve existing hardware engineers.
In this work, we look at one recent prototype system, and discuss open questions spanning from fundamental models through usable interfaces.

\end{abstract}



\maketitle

\section{Introduction}

As electronic components and circuit board fabrication have become more affordable, designing and building hardware has also become more accessible than ever before~\cite{mellis2016engaging}.
While modern board design tools involve both \textit{schematic entry} and \textit{board layout}, the schematic side has largely stagnated on graphical schematic capture, where users draw schematics by placing components on a virtual sheet and connecting their pins together.
Furthermore, our earlier formative study~\cite{ducky:beyond} found that schematic capture is only one phase -- and as currently practiced, largely data entry -- of the overall design flow, and mainstream schematic tools are unable to meaningfully support the more interesting tasks of circuit or system architecture design.

While there has been much work on improving board-level design, one recent line of academic research has been into hardware description language (HDL) approaches, as is common in digital logic design for chips and FPGAs.
While a Verilog-like approach of translating schematics into equivalent HDL offers limited benefits, a Chisel-like~\cite{izraelevitz:chisel} approach of programmatically constructing hardware via \textit{generators} can be much more powerful and re-usable.
For example, while any individual power converter subcircuit would define internal component values like capacitance and inductance specified for one application, a generator would instead define the methodology, as code, and automatically size those components for each specific application.

Furthermore, concepts borrowed from object-oriented programming, like type hierarchies and inheritance, also help raise the level of design.
Continuing the power converter example, there may be several choices of subcircuit based around different controller chips.
While modern design practice and tools require manual selection of these details, a type hierarchy of power converters could allow users to instantiate an abstract converter, and then either select from a list of compatible alternatives, or allow an automated choice.
The notion of type can also be generalized to include parameters~\cite{ramesh:edg} like voltage and current ratings, enabling correctness checks.
Ultimately, the benefits of generators and typing would serve to increase design efficiency with automation, while making design more accessible by encapsulating specialized knowledge within higher-level interfaces.

\begin{figure}
  \centering
  \begin{minipage}{.45\textwidth}
    \centering
    \begin{minted}[
      xleftmargin=18px,
      linenos,
      fontsize=\scriptsize,
      baselinestretch=1.2
    ]{python}
class Blinky(Block):
  def contents(self):
    super().contents()
    self.mcu = self.Block(MagicMcu())
    self.led = self.Block(IndicatorLed())
    self.connect(self.mcu.gnd, self.led.gnd)
    self.connect(self.mcu.digital[0], self.led.io)
    \end{minted}
    \label{fig:example_hdl}
  \end{minipage}%
  \begin{minipage}{.45\textwidth}
    \centering
    \includegraphics[scale=0.66]{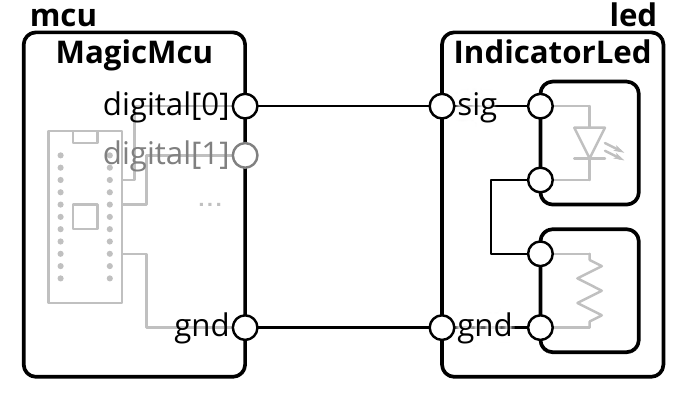}
    \label{fig:example_blocks}
  \end{minipage}
  \caption{
  Example top-level Polymorphic Blocks~\cite{ducky:polymorphic} hardware description language (HDL, left) and corresponding block diagram (right) for a device with a microcontroller and LED-resistor subcircuit.
  Lines 4 and 5 declare the two top-level blocks \texttt{mcu} and \texttt{led}, while lines 6 and 7 connect their signal and ground ports.
  The block diagram is \textit{hierarchical} in that blocks can be defined with a (sub-)block diagram, as with the \texttt{IndicatorLed} which contains an internal LED and resistor.
  The \texttt{IndicatorLed} is also written in HDL (not shown) as a generator, containing code to size the internal resistor based on its input voltage.
  }~\label{fig:example}
\end{figure}

\begin{figure}
    \centering
    \includegraphics[scale=0.66]{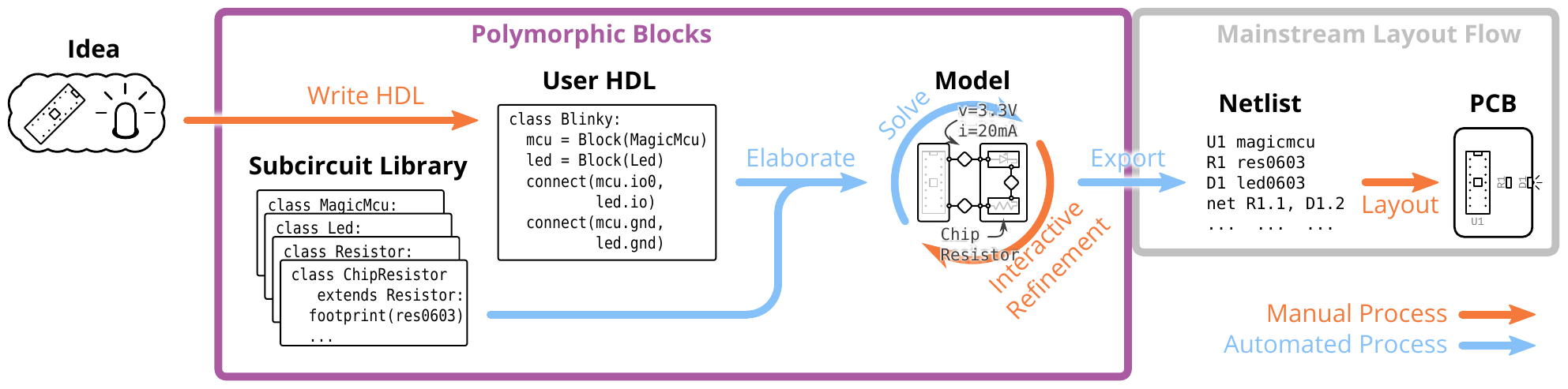}  %
    \caption{
    In the Polymorphic Blocks workflow (purple box), designers start by writing their system architecture in HDL, which is then elaborated into a hierarchy block graph model and expanded using library components.
    The system designer can further specify refinements for arbitrary blocks, replacing them with a compatible sub-type.
    With a fully expanded graph, automatically propagated parameters can be checked for correctness and consistency.
    The result can then be exported to a netlist file, which can then be imported into a board design tool for layout in mainstream tools.
    This HDL system replaces the role of schematic capture in the device design flow.
    }~\label{fig:flow}
\end{figure}

The most recent example of a board-level HDL is Polymorphic Blocks~\cite{ducky:polymorphic}, which incorporates the above concepts to support this higher-level design vision with a library-based flow.
As shown by the example in Figure~\ref{fig:example}, it is implemented as a Python-embedded domain specific language (DSL) and presents a hierarchical block diagram design model.
The design flow is shown in Figure~\ref{fig:flow}, and exports circuit data to a mainstream layout tool as part of an end-to-end board flow.

Example projects and an accompanying user study (though small) demonstrate that the ideas show promise, both in terms of design capability and usability.
Yet, we emphasize that this is only a prototype system, and user feedback also makes it clear that there is much work to be done, from devising intuitive names to fundamental design model enhancements.

In the rest of this paper, we will expand on and extend the future work suggested in Polymorphic Blocks for open avenues of research into board-level HDLs.
Areas include designing usable and powerful abstractions, developing supporting tooling, and considering larger contexts like communities.
While electrical engineering expertise might be helpful for some aspects, particularly design models, we also believe that it is not necessary for many other aspects, like usable programming environments and community building.

\section{Designing Usable Models}

\begin{figure}
    \centering
    \begin{tabular}{ l p{7cm} l }
 Level of Abstraction & Examples & Developers \\ 
 \hline
 System, Board & Any device, eg, IoT sensor, USB peripheral & Anyone (ideally) \\  
 Libraries & Resistor divider, microcontroller subcircuit,\newline buck converter subcircuit & Electrical Engineers \\  
 Abstract Blocks & Basic components: eg, resistors, capacitors;\newline abstract power converters & Electrical Engineers \\
 Electronics Model & Voltage source and sink ports, digital IO ports,\newline and checks (eg, voltage output vs limits) & Core Developers \\
 Problem Structure & Hierarchy block diagrams~\cite{ducky:polymorphic,ramesh:edg} & Core Developers \\
    \end{tabular}
    \caption{
    One possible view of the layers of abstraction for a board-level HDL, and who we expect to develop at each layer.
    The top level is meant to be the most useful and accessible, while the more infrastructural layers further down require more specialized knowledge.
    }~\label{fig:layers}
\end{figure}

Unlike mature digital logic HDLs like Verilog and VHDL, which have largely settled on abstractions, board-level HDLs are still a novel field where plenty of open questions remain.

Taking a layered view, as in Figure~\ref{fig:layers}, the fundamental design model and problem structure sits at the bottom.
Polymorphic Blocks is based on hierarchy block diagrams, which has the benefit of familiarity to existing hardware engineers while scaling across levels of abstraction.
Yet, the details reveal opportunities for improvement, for example: support for arrays of blocks and ports, defining generation order, examining advanced type constructs like union and intersection, and how these could benefit re-use and while being intuitive for users without a deep software background.
We also note that while hierarchy block diagrams have a lot of generality, other design models like behavioral~\cite{anderson:tac} and dataflow models may be more natural for certain domains.

Though the design of these models and navigating the relevant tradeoffs is more in the electrical engineering domain, there is also an important and less domain-dependent meta-point of how these underlying models can evolve.
We don't expect the first (or second, or even \textit{n}th) attempt to be comprehensive and final, yet it would also be imprudent to break backwards compatibility for every change.
While these languages are still agile, it would be important to understand what techniques are available, and what restrictions are necessary, to design for forward compatibility.

\section{Interfaces and Development Environments}

Despite the potential benefits of a board-level HDL, they are still very different than the currently dominant approach of drawing graphical schematics.
While some learning curve is inevitable, flattening it as much as possible is necessary to avoid being inaccessible.
The user feedback from Polymorphic Blocks suggests that addressing this may be critical to adoption.

Like modern programming languages, an IDE could provide a discoverable and graphical interface for working with HDL.
Extending the auto-generated block diagram visualization from Polymorphic Blocks to update live with HDL edits could help bridge the conceptual gap from text description to schematics.
Yet, such a feature also raises design questions for the HDL, such as properties needed to efficiently support live visualization, and how to render parameterized blocks independently of their instantiating environment.

A more ambitious design would go beyond visualization, and towards allowing edits from a block diagram environment.
Going full circle back to schematic capture, this system could allow a gradual transition path for current hardware engineers through familiar interfaces and a gentle onboarding path for non-programmers with a graphical interface, all while keeping the door open to the HDL.
Such users would still benefit from libraries, which have a fundamental block abstraction.
Furthermore, graphical programming languages like LabVIEW suggest that at least some generator operations can translate to graphical constructs.

We also note that there are similar trends towards wider accessibility and generators in the digital logic field, and ideas for graphical environments and tooling for board-level HDLs may also prove useful for these chip-level HDLs.

\section{Building Community}

Ultimately, programming languages serve as enablers: much of the heavy lifting in modern software development comes from the massive amount of libraries developed by large and vibrant communities.
We believe that community participation and libraries will also be crucial to realizing the full potential of a board-level HDL.

However, community libraries are a double-edged sword.
Software bugs may be tolerable in part because software is fast and cheap to update, but it costs real money and time to re-spin hardware.
As might be expected, confidence has been a recurring theme in our user studies when discussing community libraries.
While simulation and static circuit modeling approaches can help, neither is anywhere near comprehensive in modern practice.
Beyond technical approaches to correctness, we expect that community processes may be useful in bridging that gap.
Marking subcircuits ``physically fabricated and tested'' could be a simple solution grounded in current electronics practice, but additional mechanisms may be needed to establish confidence in generators instead of individual subcircuit instances.

\section{Conclusion}

By borrowing programming concepts like generator languages and type systems, board-level HDLs can both increase design efficiency with automation and make device design more accessible by encapsulating low-level knowledge in libraries.
Yet, as a nascent field, there are many open questions across the map, from models and abstractions to user interfaces.
Cross-pollination between electrical engineering, programming languages, and human computer interaction communities can help converge towards systems that are ultimately powerful \textit{and} usable.

\bibliographystyle{ACM-Reference-Format}
\bibliography{sample-base}

\end{document}